# A Green Laser Pointer Hazard

Jemellie Galang
Alessandro Restelli
Edward W. Hagley
Charles W. Clark

# NIST Technical Note 1668

# A Green Laser Pointer Hazard

Jemellie Galang
Alessandro Restelli
Edward W. Hagley
Charles W. Clark
*Electron and Optical Physics Division*
*Physics Laboratory*

July, 2010

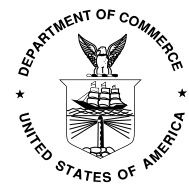

# A Green Laser Pointer Hazard

Jemellie Galang, Alessandro Restelli, Edward W. Hagley* and Charles W. Clark

*Joint Quantum Institute, National Institute of Standards and Technology and University of Maryland, Gaithersburg, MD 20899-8410*
**also Acadia Optronics, Rockville, MD 20850*

**Abstract:** An inexpensive green laser pointer was found to emit 20 mW of infrared radiation during normal use. This is potentially a serious hazard that would not be noticed by most users of such pointers. We find that this infrared emission derives from the design of the pointer, and describe a simple method of testing for infrared emissions using common household items.

## 1. Introduction

In December, 2009, we made a retail purchase of three inexpensive green laser pointers for US$15 each. These devices were advertised to have output power of 10 mW and had common packaging. One of these lasers had a much weaker green light intensity compared to the others, as judged by normal eyesight. We measured the output power of that laser and found that it emitted more than ten times more invisible infrared light than visible green light.

This is a serious hazard, since humans or animals may incur significant eye damage by exposure to invisible light before they become aware of it.

We have found that this problem is common in low-cost green laser pointers, though its seriousness varies widely. The hazards would not be immediately apparent to ordinary users. In this paper, we describe the origin of this problem, and we suggest a simple method for diagnosing it that can be used by those who have access to a digital or cell phone camera and an inexpensive web camera.

## 2. Principles of inexpensive green laser pointer operation

The common green laser pointer is a technological marvel that has recently become a commodity item. Twenty years ago, 10 mW of coherent green light could be obtained only from research laboratory-grade lasers that cost over $100,000 and occupied much of an experimental laser table. Today's green laser pointer delivers much the same performance for less than $20, in a package the size of a slim cigar, powered by two AAA batteries.

The atomic energy level diagram of the green laser pointer is shown in Fig. 1. Hereafter, we use the acronym GLP (Green Laser Pointer) to refer only to devices employing the diode-pumped, frequency-doubling architecture shown there. Figure 1 exhibits three essential elements of operation, each of which was at its time a highlight of laser technology:
- a semiconductor pump diode laser operating at an infrared wavelength of 808 nm
- a  neodymium ion oscillator at a wavelength of 1064 nm
- a frequency-doubling crystal that generates light of half of that wavelength, the familiar GLP color at 532 nm

Today, all three of these functions are integrated in an inexpensive package that is produced in high-volume manufacturing. A schematic of a typical GLP is shown in Fig. 2.

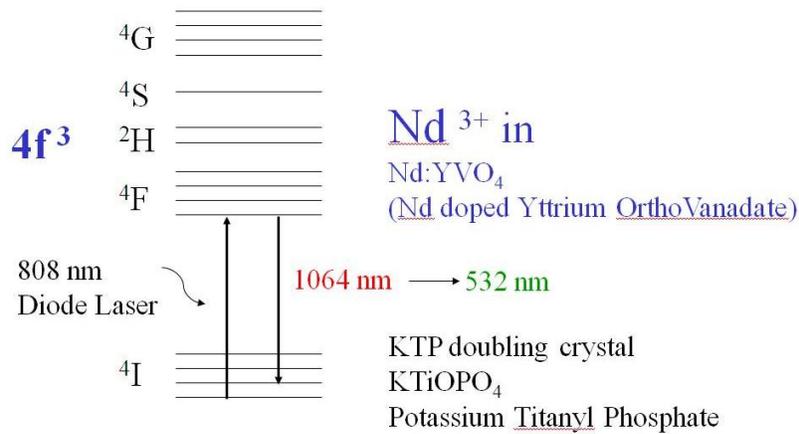

**Figure 1.** Atomic energy levels and light involved in the operation of a GLP. Triply-charged ions of the neodymium atoms, $Nd^{3+}$, are present as dopants in a crystal of yttrium orthovanadate ($Nd:YVO_4$). The $Nd^{3+}$ ion contains three 4f electrons outside closed electron shells, so its electronic configuration is thus designated $4f^3$. The horizontal lines indicate some of the energy levels of the $4f^3$ configuration, labeled by conventional spectroscopic notation. A semiconductor diode laser with an infrared wavelength of 808 nm excites the lowest $^4I$ state to an electronically excited $^4F$ state. The $Nd^{3+}$ ion emits infrared radiation, at a wavelength of 1064 nm, as it drops from the excited state into a different $^4I$ state. This radiation is directed into a "frequency doubling" crystal of potassium titanyl phosphate, which emits light of half the wavelength: 532 nm. Figure credit: Dr. Joseph Reader, National Institute of Standards and Technology.

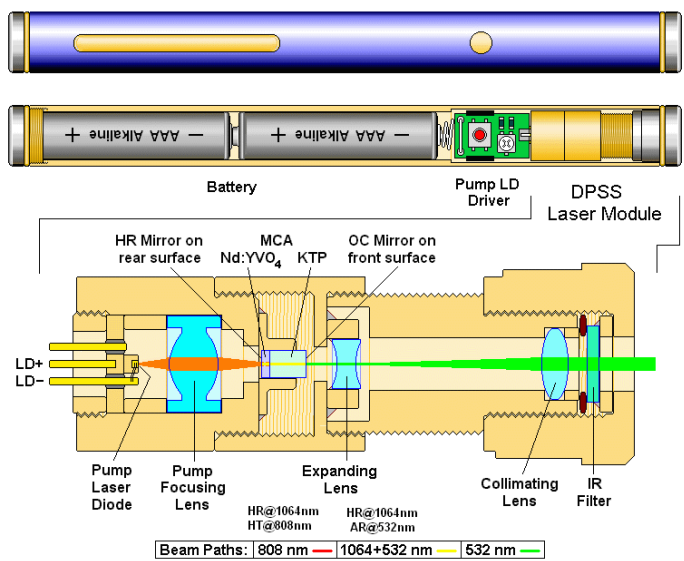

**Figure 2.** Schematic of the operation of a GLP based on a Multiple Crystal Assembly (MCA). The familiar external package contains two AAA batteries that power the unit, a printed circuit board with the pump Laser Diode (LD) driving circuitry and the Diode Pumped Solid State Laser (DPSS) Module. The 808 nm pump LD is optically coupled to the $Nd:YVO_4$ conversion crystal (violet section of the MCA), which emits 1064 nm light into the KTP frequency-doubling crystal (light blue section). The 532 nm light from the KTP crystal is sent through an expanding and collimating lens assembly to produce a collimated output beam. In this configuration, an IR Filter prevents the 808 nm and 1064 nm light from exiting the laser. In the GLP discussed in this paper, no such infrared filter was present. HR: High Reflectivity; HT: High Transmissivity; AR: Anti-Reflective; OC: Output Coupler; IR: InfraRed. This figure is the copyrighted work of Samuel M. Goldwasser [1], and is reprinted here with his permission.

## 3. Standard hazards of GLP operation

A 10 mW GLP is rated as a Class IIIb device [2], which is capable of causing serious eye damage. However, there is little documented data on eye injuries actually caused by exposure to GLPs. The light from a GLP is at a wavelength to which the human eye is most sensitive (Fig. 3). This is why a GLP appears far brighter than a red laser pointer that emits the same optical power. Thus, GLPs are superior to red laser pointers for demonstration purposes. Direct exposure of the eye to GLPs seems to be relatively rare, because of the blink reflex that protects the eye from light of strong visual intensity.

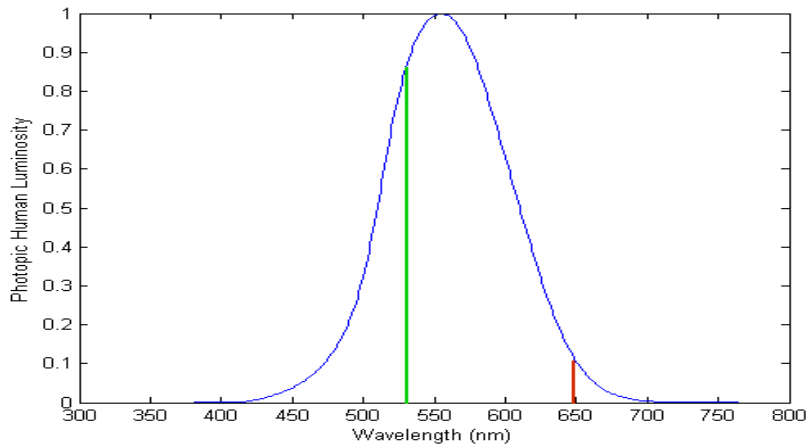

**Figure 3.** Diagram of human visual response showing the wavelength of the GLP line at 532 nm, and the 650 nm wavelength of a typical red laser pointer. The blue curve shows the perceived brightness of light from a source of uniform intensity across the wavelength range. Thus a 5 mW GLP would appear as bright as a 41 mW red laser. Data from the 1931 report of the International Commission on Illumination (CIE), as presented in Ref. [3]

## 4. Less-appreciated hazards of the GLP

The green light of a GLP is produced by much more intense sources of invisible infrared light: the diode pump laser and the neodymium oscillator. In some green laser pointer designs, this invisible radiation is blocked from the output beam by an infrared-blocking filter, as shown in Fig. 2. In the case that we report here, the infrared radiation was not blocked, due to the absence of this filter. Intense infrared radiation poses a grave danger to humans and animals because, being invisible, it does not activate the blink reflex. Thus, a considerable dose of infrared radiation can enter the eye and cause retinal damage, imperceptible to the victim before serious damage occurs. Invisible lasers, in the infrared and ultraviolet, are among the most frequently reported sources of laser-induced eye injuries, with many reported instances of permanent damage to vision [4]. A survey of the 100 accidental, non-medical laser-induced eye injuries that had been reported in the scientific literature up to 1999 found that Nd-based lasers with wavelengths of 1060-1064 nm were implicated in 49 % of all cases, while 532 nm lasers accounted for only 7 %. [5]

The main danger associated with GLPs occurs in cases of low conversion efficiency from infrared to visible radiation. This can result from the manufacturing process. Normally, if the conversion efficiency is high, the inter-cavity infrared power will be low because conversion to green light draws down the power of the infrared. However if the conversion efficiency is low, then the inter-cavity infrared power can build up to high levels, resulting in strong infrared emission. In the extreme case of zero conversion efficiency, it would be possible for the GLP to emit intense infrared light but no visible green light. Inclusion of an infrared-blocking filter in GLP design can prevent such infrared emission. As shown in Fig. 4, for the laser that we discuss here, about 10 times more light was emitted in the infrared than in the green. This was due to both low infrared/green conversion efficiency and the absence of an infrared-blocking filter. The total emitted power as measured with a thermal detector was 20 mW. It is dangerous for such intensities of infrared laser light to be present in an uncontrolled environment. For example, many windows in modern commercial and industrial buildings have infrared reflective coatings to reduce solar heating of the interior. If a green laser is directed on such a window, the green light will pass through it, but any infrared emitted by the GLP will be reflected. This is a substantial hazard. Any differential reflection of infrared vs. green light poses such hazards, since the green light that travels with the infrared light may not be sufficiently intense to activate the blink reflex.

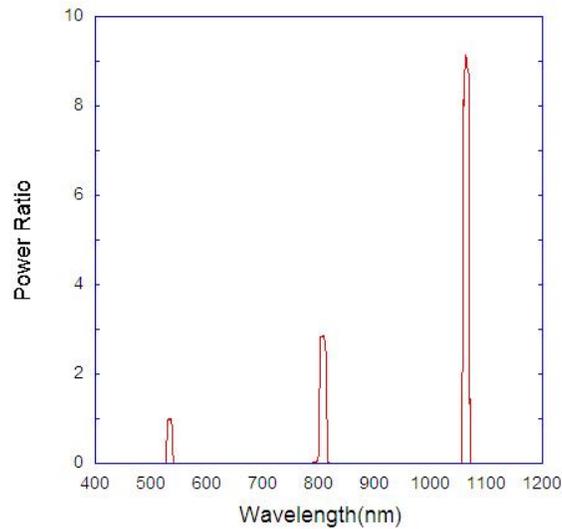

**Figure 4.** Power spectrum of the laser under test, showing the ratios of the measured power of each laser line to the measured power of the green line at 532 nm.

## 5. A simple test for green laser pointers

Inexpensive green laser pointers are produced in vast quantities. Few that we have seen carry traceable manufacturers' marks that are visible to the consumer. Widespread production of this commodity item is likely accompanied by inconsistent adherence to practices of safe design and/or quality control. The drive to lower costs on such commodity items may encourage manufacturers to eliminate the infrared-blocking filter, resulting in the type of hazard described here, especially when the conversion efficiency from infrared to green is low. The consequence has been the apparent release into the U.S. retail market of GLPs that emit potentially hazardous and unappreciated levels of infrared radiation.

We now describe a simple test that can be used to detect infrared leakage which can be performed by anyone with access to a common digital or cell phone camera, a compact disc (CD), a web camera and a TV remote control. This test is an elementary exercise in laser spectroscopy, which uses a diffraction grating to separate light of different wavelengths onto a viewing screen [6]. Despite the simple nature of the test, we cannot over-emphasize the importance of following proper laser safety procedures in carrying it out [4].

**CAUTION**

**DIRECT EXPOSURE ON THE EYE BY A BEAM OF LASER LIGHT SHOULD *ALWAYS* BE AVOIDED WITH ANY LASER, *NO MATTER HOW LOW THE POWER***

**NEVER LOOK INTO A DIRECT, REFLECTED OR DIFFRACTED LASER SOURCE**

**TEST PROCEDURE MAY PRODUCE VISIBLE AND INVISIBLE LASER RADIATION**

**KEEP ALL LASER BEAMS AT TABLE LEVEL; KEEP EYES WELL ABOVE TABLE LEVEL**

**WEAR SAFETY GLASSES APPROVED FOR LASER AND INFRARED RADIATION WHEN CONDUCTING TEST PROCEDURE**

**CONDUCT TEST PROCEDURE ONLY UNDER PROPER SUPERVISION IN A PROPERLY CONTROLLED ENVIRONMENT**

A simple layout of this experiment is shown in Fig. 5. The CD serves as the diffraction grating, which deflects light of different wavelengths in different directions. The digital or cell phone camera is sensitive only to visible light, and is used to photograph the green diffraction pattern, as shown in Fig. 6. On the other hand, inexpensive web cameras are, or may easily be modified to be, sensitive to both visible and infrared radiation. An infrared TV remote control unit may be used to determine whether the web camera is infrared sensitive; if it is not, it can be modified by the procedure described in Appendix B. The infrared-sensitive web camera is then used to photograph the diffraction pattern. Comparison of images acquired by digital and web cameras determines the presence of infrared light, as shown in Fig. 6 below.

## 6. Details of test procedure

We now describe an actual implementation of this experiment in more detail.

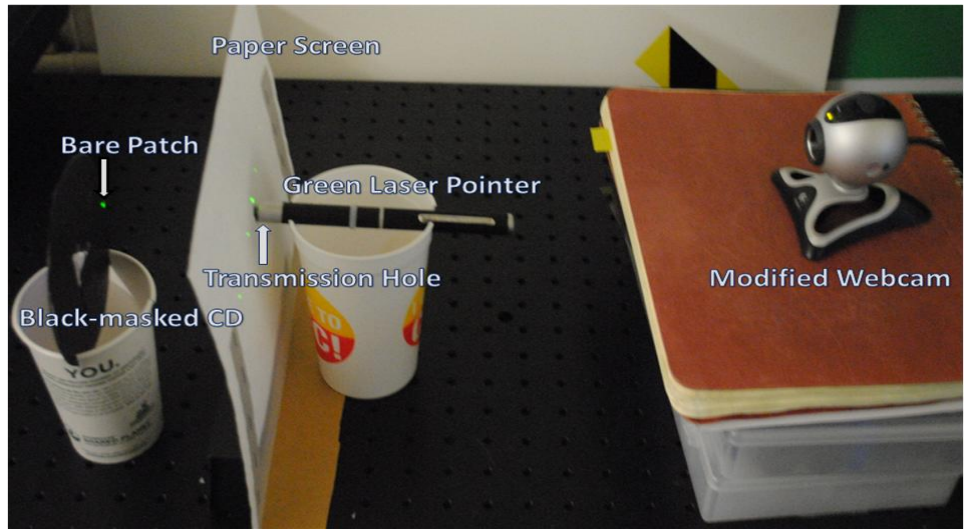

**Figure 5.** Experimental setup for determining whether infrared radiation is emitted by a GLP. The GLP light passes through a hole in the paper screen and is diffracted by the black-masked CD on the left. Five green diffraction spots are visible on the paper screen.

Ordinary disposable drink cups (or stacks of books) are used as optical mounts for the GLP, the CD, and the cameras.

- Pressing a knife or scissors blade across the diameter of a cup cuts slots that provide an adjustable vertical mount for the CD.

- Another cup provides a level mount for the laser, whose power button can be held on by duct tape, a binder clip, or a cable tie.

- A small transmission hole is made in a piece of ordinary ink-jet or copier paper, which is positioned between the laser and the CD. The transmission hole allows the laser light to pass through the paper and hit the CD. Black tape is used to cover most of the CD surface, leaving a small bare patch exposed to the laser spot. This suppresses distracting reflections of the light from other parts of the CD surface.

- The light reflected from the bare patch on the CD returns to the paper, which forms a back-lit screen that can be viewed by the eye or photographed from behind the laser. The pattern of visible light in this configuration is shown in the top frame of Fig. 6. The center spot is the direct reflection of the laser by the CD. The spots on either side are the diffraction orders described in Appendix A below.

- After viewing and understanding this pattern, and photographing it with a digital camera, we place an infrared-sensitive web camera so as to view the same scene. Such infrared-enabled web cameras will show the presence of infrared light, as shown in the bottom frame of Fig. 6.

## 7. Origin of the problem

After studying the visible and infrared light emitted by this GLP, we disassembled it. No infrared-blocking filter of the type shown in the schematic of Fig. 2 was found in this particular device. Inspection of the assembly revealed that the manufacturer had not incorporated a holder for such a filter. We thus believe that the absence of the filter in this case was due to a design decision.

## 8. Summary

We describe a green laser pointer purchased in ordinary commerce which could potentially expose the eye to dangerous levels of infrared radiation. We devised a procedure by which infrared radiation from a green laser pointer may be detected with common household apparatus.

## Appendix A. Image details

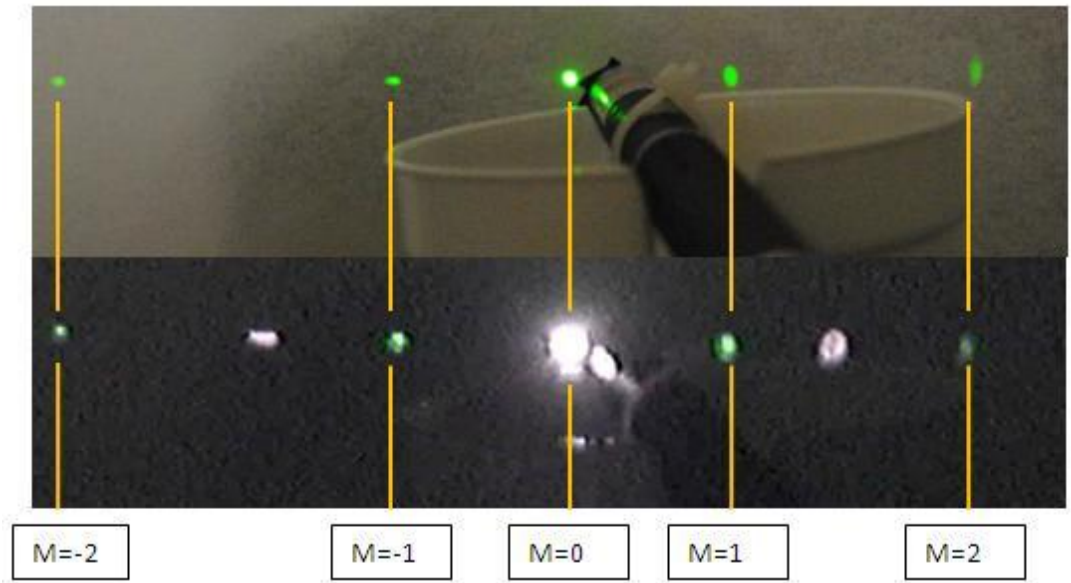

**Figure 6.** Top frame: ordinary camera photograph of diffraction of green laser light by the masked CD, from the vantage point of the web camera shown in Fig. 5. The GLP is visible in the foreground. The yellow lines are aligned with the green diffraction spots corresponding to zero-, first- and second-order diffraction. Bottom frame: the same vignette, photographed by a web camera with no infrared-blocking filter. Here we see intense diffraction spots that are not visible to the eye, between the first- and second-order diffraction spots of the green light. These are due to 808 nm infrared light. Note that the infrared spots exhibit greater divergence than the green. The infrared light spreads out beyond the green, which could be injurious, for example, to a cat closely chasing a spot of green light.

The CD acts as a diffraction grating with a grating spacing defined by the CD tracks that are separated by $d = 1.6$ μm. Knowing the grating separation $d$, it is possible to calculate the angle of diffraction, $\theta$, as a function of wavelength $\lambda$. For normal incidence it is possible to use the simple equation [7]

$$\sin \theta = m\lambda/d$$

where $m = 0, \pm1, \pm2$, etc. is an integer. The formula indicates that each wavelength will be scattered back in a discrete series of angles that match the integral values of $m$. The case $m = 0$ corresponds to simple back reflection.

The two frames of Fig. 6 show images taken with digital (visible) and web (visible + infrared) cameras. The top frame was acquired with an ordinary digital camera, which has a high-quality infrared-blocking filter. It corresponds closely to what would be seen by the naked eye. Only the green light is seen, distributed left to right among diffraction orders corresponding to $m = -2, -1, 0, 1$ and $2$, as labeled.

The bottom frame was acquired by an inexpensive web camera, with its infrared-blocking filter removed, as described in Appendix B. Here we see the 808 nm infrared light, with a central ($m = 0$) spot much more intense than the green, and with bright $m = \pm1$ spots outside the fainter $m = \pm1$ spots of the green, as predicted by the diffraction equation above.

The webcam sensitivity cuts off near wavelengths of 1064 nm, so we cannot see the diffraction spots ($m = \pm1$) corresponding to this wavelength. If detectable, such spots would overlap the $m = \pm 2$ spots of the 532 nm green light.

**Appendix B. Using an inexpensive web camera as an infrared detector**

As mentioned in Section 6 above, not all web cameras are sensitive to infrared light. Web cameras that are sensitive to infrared are usually labeled as night vision cameras. Not capable of imaging in complete darkness, such cameras use an infrared light-emitting diode to illuminate the subject. It is possible to modify most other web cameras to extend their sensitivity into the infrared range. There is a vast internet-based literature on such modifications, which can easily be found by submitting "IR webcam" to a search engine. For completeness, we describe the actual procedure that we used ourselves to modify such a camera. However, we do not claim it to be optimal or even recommended. It is similar to the procedures that are described online.

After disconnecting the web camera from all data communication and power lines, we disassemble the housing by unfastening screws. Figure 7 shows the interior components of the camera. The infrared-blocking filter is mounted close to the camera's sensor, and while appearing transparent it reflects red light if observed laterally.

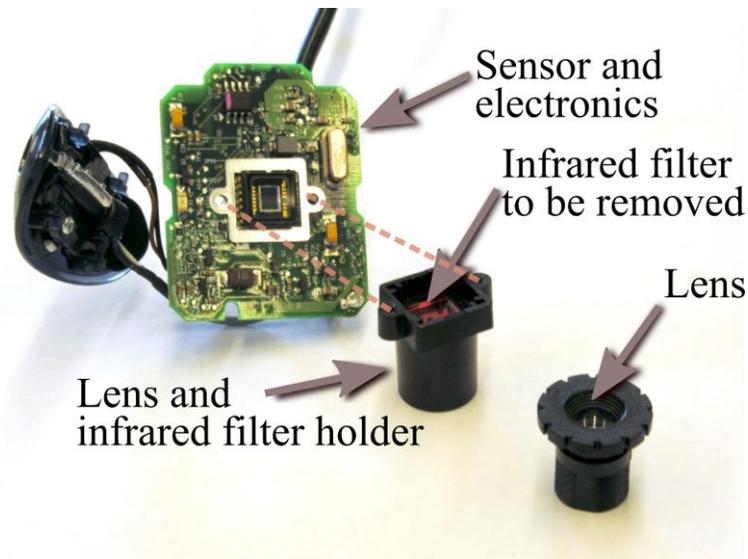

**Figure 7.** Image of a web camera that was disassembled in order to remove the infrared-blocking filter that is usually located in proximity to the camera's sensor. In this particular model the filter is glued inside the lens holder. The holder is connected to the main circuit board using two screws through mounting holes indicated by the dashed lines. Lens and holder are threaded for both mechanical assembly and adjustment of focus.

Since the procedure of filter removal involves access to electrical and mechanical components, we recommend this operation only to those familiar with delicate disassembly and assembly work. Power and data cables must be disconnected during the modification, and the camera package should be fully reassembled before subsequent operation. This procedure is likely to void all manufacturers' warranties.